\documentclass[]{aa}  

\usepackage{booktabs}
\usepackage{graphicx}
\usepackage{subcaption}
\usepackage{tikz}
\usepackage{tabularx}
\usepackage{multirow}
\usepackage{threeparttable}
\usepackage{ragged2e} 
\usepackage{chemformula}
\usepackage{amsmath}
\usepackage{placeins}
\usetikzlibrary{shapes.geometric, arrows.meta, positioning}

\newcommand\rev[1]{\textcolor{black}{#1}}
\begin{document}

\title{Surrogate-Accelerated Bayesian Inversion for Exoplanet Interior Characterization}

\author{T. de Wringer\inst{1}
\and C. Dorn\inst{2}\fnmsep\thanks{Corresponding author: dornc@ethz.ch}
\and E. O. Garvin\inst{2}
\and S. Marelli\inst{3}
}

\institute{
Seminar für Statistik, ETH Zürich, Raemistrasse 101, 8092 Zürich, Switzerland
\and Institute for Particle Physics and Astrophysics, ETH Zürich, Wolfgang-Pauli-Strasse 27, 8093 Zürich, Switzerland
\and Department of Civil, Environmental and Geomatic Engineering, ETH Zürich, Stefano-Franscini-Platz 5, 8093 Zürich, Switzerland
}

\date{Accepted December 16, 2025}

\abstract
{Characterizing the interior structure of exoplanets is an inverse problem often solved using Bayesian inference, but this approach is hampered by the high computational cost of planetary structure models. To overcome this barrier, we present a robust framework that accelerates inference by replacing the computationally expensive physics-based forward model with a fast polynomial chaos-Kriging (PCK) surrogate directly within a Markov chain Monte Carlo (MCMC) sampling loop. We rigorously validate our approach using a suite of tests, including a direct comparison against a benchmark MCMC inference using the full forward model, and a large-scale coverage study with 1000 synthetic test cases to demonstrate the statistical reliability of our inferred credible intervals. Our surrogate-assisted framework achieves a computational speedup of over 2 orders of magnitude (factor of $\sim$320), reducing single-CPU inference times from days to minutes. This efficiency is achieved with a surrogate that requires only a few hundred forward model evaluations for training \rev{for a single planet}. This data efficiency provides significant flexibility for model developments and a clear advantage over common machine learning approaches, which typically demand vast training sets ($>10^6$ model runs) and intensive pre-computation. The PCK surrogate maintains high fidelity with $R^2 > 0.99$ for most scenarios, and root-mean-square errors typically an order of magnitude smaller than observational uncertainties. This efficiency enables large scale population studies while preserving statistical robustness, which is computationally impractical with traditional methods.}


   \maketitle
   \nolinenumbers
\section{Introduction} \label{sec:introduction}
Understanding the interior structure of exoplanets is fundamental to planetary science, as it provides crucial insights into planetary formation processes, habitability potential, and the diversity of worlds beyond our solar system. With over 6,000 confirmed exoplanets discovered to date, the field has moved beyond simple detection toward detailed characterization, seeking to understand not just what these planets are like on the surface, but what lies beneath.

Traditional approaches to exoplanet interior characterization rely on Bayesian inverse modeling, where observable quantities like mass and radius are used to infer internal structure parameters such as core size, mantle composition, and atmospheric properties. This process requires a forward model, $f(\theta)$, which maps interior structure parameters $\theta$ (e.g., core and mantle mass fractions) to observable properties. The inverse problem then seeks to constrain the parameter space of $\theta$ by employing Bayesian inference frameworks, typically using Markov Chain Monte Carlo (MCMC) methods to sample the posterior probability distribution of the interior parameters \citep{dorn2025interior}.

However, this approach faces a fundamental computational bottleneck. Recent developments in interior modeling have led to increasingly sophisticated physics-based forward models that \citep[incorporate thermodynamically stable mineralogy, volatile dissolution, and evolved compositions;][]{dorn2021hidden, luo2024interior, rogers2025most}. While these advances improve physical realism, they come at a steep computational cost, with individual forward model evaluations requiring seconds to minutes of computation time. Since MCMC methods typically require tens or hundreds of thousands of forward model evaluations to ensure convergence, direct application becomes computationally impractical, with single-planet characterizations taking days of computation time \citep{Haldemann2023}.

This computational barrier poses a challenge for exoplanet science. It limits the scope of studies that can be undertaken, prevents large-scale population analyses, and restricts the statistical rigor that can be applied to individual characterizations. The challenge is particularly acute given the exponentially growing catalog of confirmed exoplanets and the wealth of data expected from upcoming missions. Moreover, the inherent degeneracy in interior structure inference—where very different internal compositions can produce the same mass and radius measurements \citep{Rogers_2010}, leading to weak constraints on planetary properties \citep{otegi2020impact}. Consequently, overcoming this uncertainty requires exploring many potential outcomes, a task that greatly increases the computational demand.

Recent work has focused on using machine learning to bypass the computationally expensive MCMC process entirely. For example, \citet{Baumeister2023} introduced ExoMDN, a method based on Mixture Density Networks (MDNs), to directly predict the posterior probability distribution of exoplanet interior structures from observables. Similarly, \citet{Haldemann2023} employed conditional Invertible Neural Networks (cINNs) for the same task. Although these approaches can replace the MCMC sampler, \rev{as done in \citep{Egger}}, their reliance on large training datasets ($10^5$–$10^6$ evaluations) substantially limits their adaptability to any developments in the underlying model, which would require a computationally expensive reconstruction of the training data. \rev{This issue is exemplified by \citep{Egger}, where a general-purpose dataset of five million samples was generated (80\% for training purpose).}

To overcome these limitations, we present a fundamentally different approach: a surrogate-accelerated Bayesian inference framework that replaces computationally expensive physics-based forward models with fast, high-fidelity approximations directly within the MCMC sampling loop. Our method employs polynomial chaos-Kriging (PCK) surrogates \citep{Schobi2015PCK,Kersaudy2015JCP}, a hybrid technique that combines the global approximation strength of sparse polynomial chaos expansions \citep{Luthen2021SIAM} with the local interpolation accuracy of Gaussian Process modelling \citep{Sacks1989Kriging}. Crucially, these surrogates can be trained using only a relatively small number of forward model evaluations (typically $O(10^{1-2})$), making them highly data-efficient compared to the machine learning approaches described above. \rev{We initially tested neural networks as potential surrogate models, but ultimately excluded them because they require substantially more hyperparameter tuning, perform poorly on relatively small training sets, and do not provide reliable uncertainty quantification. By contrast, PCKs are theoretically better suited for case-by-case training and offer stable performance together with principled uncertainty estimates. In the following, we demonstrate the practical implications and performance of this framework.}

A gain in computational speed introduces a new source of uncertainty that must be carefully managed: model discrepancy. When replacing a slow but precise planetary structure model with a fast surrogate, we inevitably introduce approximation errors. This discrepancy is often non-uniform across parameter space; its magnitude can depend complexly on the input parameters. Ignoring this source of uncertainty can lead to systematically biased and overconfident conclusions about an exoplanet's interior \citep{Kennedy2001}. Our framework addresses this challenge by quantifying and propagating surrogate uncertainty through the inference process.

Our approach addresses several critical challenges in the field while preserving the robust statistical guarantees of traditional MCMC methods. First, it reduces single-CPU inference times from days to minutes while maintaining statistical rigor and physical fidelity. Second, unlike inflexible neural network approaches that require complete retraining when physical models evolve, our surrogate-based method can rapidly adapt to model updates with minimal computational overhead. Third, it enables \rev{substantially more efficient} large-scale studies and comprehensive validation analyses that are essential for understanding exoplanet population diversity.

In this work, we rigorously validate our approach through multiple complementary analyses, including direct comparison against benchmark MCMC runs using full forward models, and a comprehensive coverage study with 1,000 synthetic test cases to demonstrate statistical reliability. The framework is demonstrated across diverse planetary scenarios, from Earth-like worlds to sub-Neptunes, and applied to the real exoplanet TOI-270 d.

The remainder of this paper is organized as follows: Section \ref{sec:methods} describes our methodology, including the forward model, surrogate construction, and Bayesian inference framework; Section \ref{sec:data} details the synthetic and real datasets used for validation; Section \ref{sec:results} presents our results across multiple validation stages; Section \ref{sec:discussion} discusses the implications and future developments; and Section \ref{sec:conclusion} summarizes our conclusions.

\section{Methods} \label{sec:methods}
\begin{figure*}[htbp]
    \centering
    
    \tikzset{
        block/.style={rectangle, draw, fill=blue!10, text width=6em, text centered, rounded corners, minimum height=4em},
        line/.style={draw, -{Stealth[length=3mm]}},
        process/.style={midway, font=\small}
    }
    
    \begin{tikzpicture}[node distance = 1.5cm and 2.7cm]
        \node[block] (params) {Prior Parameter Space ($\theta$)};
        \node[block, right=of params] (forward) {Slow Forward Model};
        \node[block, text width=8em, right=of forward] (dataset) {Training Dataset, pairs of $\{\theta_i, y_i\}$}; 
        \node[block, text width=7em, right=of dataset] (surrogate) {Fast, Trained Surrogate Model ($\hat{y}(\theta)$)};
        
        \node at (current bounding box.north) [above=0.2cm] {(a) One-Time Training Phase};

        \path [line] (params) -- node[process, above] {Run $N_{\text{train}}$ times} (forward);
        \path [line] (forward) -- node[process, above] {Generates} (dataset); 
        \path [line] (dataset) -- node[process, above] {Trains} (surrogate); 
    \end{tikzpicture}
    
    \vspace{1.5cm} 
    
    \begin{tikzpicture}[node distance = 1.5cm and 2cm]
        \tikzset{mcmcblock/.style={rectangle, draw, fill=green!10, text width=8em, text centered, rounded corners, minimum height=5em}}

        \node[mcmcblock] (mcmc) {MCMC Sampler};
        \node[block, text width=7em, left=of mcmc] (obs) {Observational Data ($y_{\text{obs}}, \sigma_{\text{obs}}$)};
        \node[block, text width=7em, below=of obs] (prior) {Prior Distribution ($P(\theta)$)};
        \node[block, right=of mcmc] (posterior) {Posterior Distribution ($P(\theta|y_{\text{obs}})$)};
        \node[block, text width=7em, below=of mcmc] (surrogate) {Fast, Trained Surrogate Model ($\hat{y}(\theta)$)};
        
        \node at (current bounding box.north) [above=0.2cm] {(b) Rapid Inference Phase};

        \path [line] (obs) -- (mcmc);
        \path [line] (prior) -- (mcmc);
        \path [line] (mcmc) -- (posterior);
        \path [line] (surrogate) -- node[process, right] {Used for Likelihood} (mcmc);
    \end{tikzpicture}
    
    \caption{A schematic of our surrogate-assisted inversion framework. 
    \textbf{(a) One-Time Training Phase:} A computationally expensive forward model is run $N_{\text{train}}$ times to generate a training database. This database is used to train a fast Polynomial Chaos-Kriging (PCK) surrogate model. This is a one-time computational cost. 
    \textbf{(b) Rapid Inference Phase:} For a given exoplanet with observational data ($y_{\text{obs}}$), the fast surrogate is placed inside an MCMC sampling loop to efficiently explore the posterior distribution of the interior structure parameters ($\theta$).}
    \label{fig:workflow}
\end{figure*}
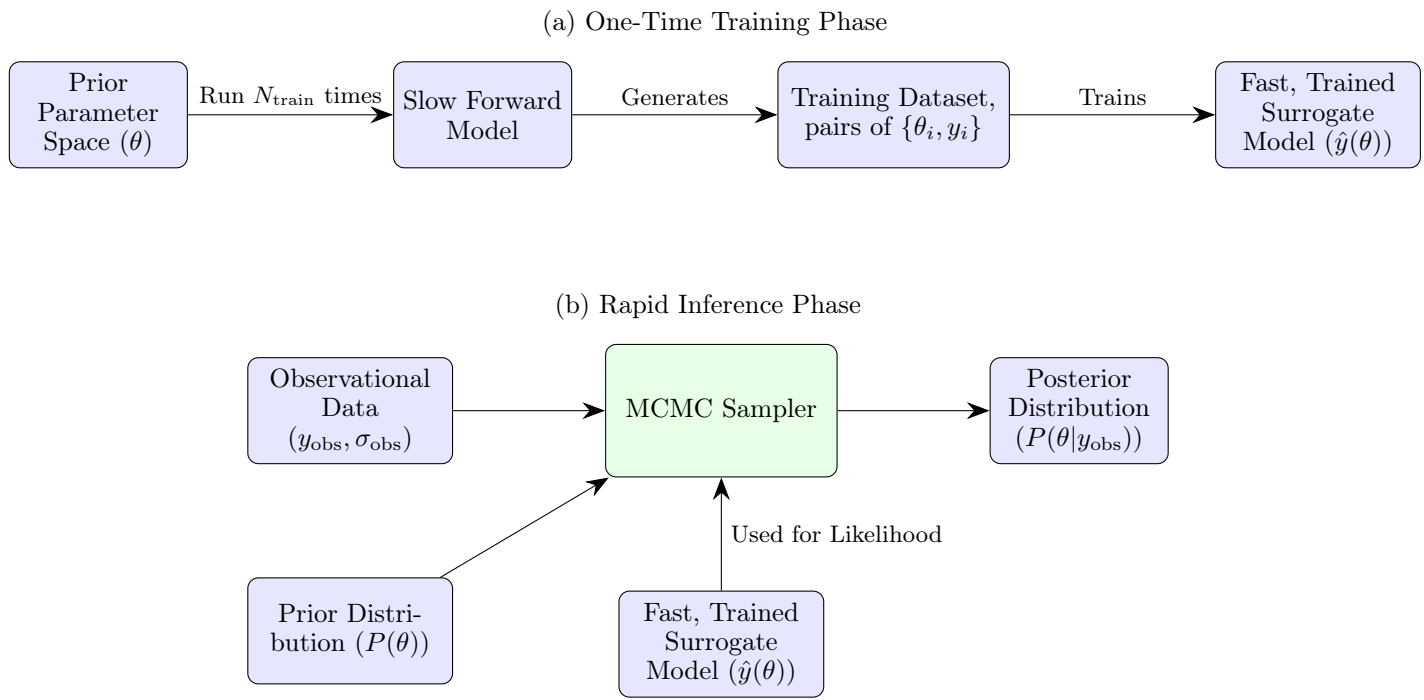

Our method for characterizing exoplanet interiors is a two-phase framework, illustrated in Figure~\ref{fig:workflow}. The first is a one-time training phase, which is an upfront computationally intensive step to build a fast surrogate model. The second is a rapid inference phase, where this surrogate is used to efficiently characterize the interior of a specific planet.

\subsection{Forward Model}
\label{sec:forward_model}
The forward model calculates a planet's radius for a given mass and composition by solving the hydrostatic equilibrium equation coupled with a thermodynamic approach. The model is based on \citet{Dorn2015, dorn2017generalized} and recent model updates allowing for miscible phases \citep{dorn_hidden_2021,luo2024interior} were not actively used here.

The model consists of three layers: an iron core, a silicate mantle, and a H$_2$-He-H$_2$O atmosphere. We assume an adiabatic temperature profile for the core and mantle and allow for both liquid and solid phases in the two layers.
For liquid iron we use the equation of state (EOS) by \citet{luo2024interior}. For solid iron, we use the EOS for hexagonal close packed iron \citep{hakim_new_2018,miozzi_new_2020}. For pressures below $\approx 125\,$GPa, the solid mantle mineralogy is modelled using the thermodynamical model \textsc{Perple\_X} \citep{connolly_geodynamic_2009} considering the system of MgO, SiO$_{2}$, and FeO. 
At higher pressures we define the stable minerals \textit{a priori} and use their respective EOS from various sources \citep{hemley_constraints_1992,fischer_equation_2011,faik_equation_2018,musella_physical_2019}. The liquid mantle is modelled as a mixture of Mg$_2$SiO$_4$, SiO$_2$ and FeO \citep{melosh_hydrocode_2007,faik_equation_2018,ichikawa_ab_2020,stewart_shock_2020}, and mixed using the additive volume law. 

The H$_2$-He-H$_2$O atmosphere layer is modelled using the analytic description of \citet{guillot_radiative_2010} and consists of an irradiated layer on top of a non-irradiated layer in radiative-convective equilibrium. \rev{The planet intrinsic luminosity is calculated following the luminosity model of \citep{mordasini_planetary_2020} and is a function of planet mass, atmospheric mass fraction and planet age.}
The water mass fraction in the envelope and atmosphere is given by the metallicity $Z_{env}$ and the hydrogen-helium ratio is set to solar.
The two components of the atmosphere, H$_2$/He and H$_2$O, are again mixed following the additive volume law and using the EOS by \citet{chabrier2019new} for H$_2$/He and the ANOES EOS \citep{1990_thompson_aneos} for H$_2$O. \rev{In \citep{guillot_radiative_2010}, visible and thermal opacity and thus their ratio $\gamma = \kappa_\mathrm{v}/\kappa_\mathrm{th}$ are treated as free parameters. To reduce this freedom, we adopt the calibration of \citet{jin_planetary_2014}, who provide $\gamma$ as a function of equilibrium temperature based on more detailed radiative–equilibrium models with wavelength-dependent opacities \citep[as done in][]{dorn2017generalized}. Implementing this calibration allows us to capture the essential physics of atmospheric absorption and re-irradiation without performing full opacity calculations. The transit radius is evaluated where the chord optical depth is $\tau_\text{ch}=0.56$ \citep{2008Lecavelier}.}

While this physics-based approach provides high-fidelity results, it comes with significant computational costs. A single evaluation of the forward model requires approximately 1.5 seconds of computation time with an Apple M4 processor and 16 GB of RAM. Consequently, a typical 100,000-sample MCMC run to determine the interior of a single planet would take roughly 42 hours, making direct MCMC sampling computationally impractical.

\subsection{Surrogate Modeling Framework}
\label{sec:surrogate_modeling}
To mitigate the computational burden of the forward model within our MCMC sampler (Fig.~\ref{fig:workflow}a), we construct a surrogate model using polynomial chaos-Kriging (PCK, \cite{Schobi2015PCK}). The PCK model is a form of universal Kriging where a sparse polynomial basis serves as the global trend, while a Gaussian process captures the remaining local variations in the model response.

Our implementation utilizes the sequential PC-Kriging (SPCK, \cite{Schobi2015PCK}) algorithm available in the UQLab software library \citep{Marelli2014UQLab}\footnote{For compatibility with the rest of our computational workflow, we adopted the python client of the cloud-based version of UQLab, UQCloud (\cite{marelli2024uqpylab}).}. 
Sequential PC-Kriging is a two-step algorithm that first calculates a sparse, basis adaptive polynomial chaos expansion on the training data using least-angle regression (LARS, \cite{Efron2004LAR,Blatman_2011}). 
This both identifies a set of polynomials orthogonal to the input distributions, and orders them by their relative contribution to the response variance.
In a second step, it fits a sequence of universal Kriging models to the training data, each using an increasing number of the previously selected polynomial basis functions as their trend. 
The accuracy of each Kriging model is assessed via cross-validation on the training data, and the best performing one is returned as the final PCK model.

The entire surrogate is trained using samples generated by the full forward model and assessed against a held-out validation set to ensure predictive accuracy. While generating the training database of several hundred samples takes approximately 15-30 minutes, fitting the surrogate model itself is highly efficient, typically completing in just minutes.

\subsection{Bayesian Inference Framework}
\subsubsection{Bayesian Statistical Model}
We infer the interior structure parameters, $\boldsymbol{\theta}$, by sampling the posterior probability distribution given the observed data, $\mathbf{y}_{\text{obs}} = \{M_{\text{obs}}, R_{\text{obs}}\}$. Following Bayes' theorem:
\begin{equation}
P(\boldsymbol{\theta}|\mathbf{y}_{\text{obs}}) \propto \mathcal{L}(\mathbf{y}_{\text{obs}}|\boldsymbol{\theta}) \cdot P(\boldsymbol{\theta}), \label{eq:bayes}
\end{equation}
where $P(\boldsymbol{\theta})$ represents the prior distribution and $\mathcal{L}(\mathbf{y}_{\text{obs}}|\boldsymbol{\theta})$ is the likelihood function.

The prior distribution $P(\boldsymbol{\theta})$ encodes the existing knowledge of and assumptions about the interior structure parameters. The likelihood $\mathcal{L}(\mathbf{y}_{\text{obs}}|\boldsymbol{\theta})$ assumes that the observational errors for mass and radius are independent and follow Gaussian distributions as commonly assumed. The total likelihood is therefore the product of the individual normal probability densities:
\begin{equation}
\mathcal{L}(\mathbf{y}_{\text{obs}}|\boldsymbol{\theta}) = \prod_{i \in \{M,R\}} \mathcal{N}(y_{i,\text{obs}} | f_i(\boldsymbol{\theta}), \sigma^2_{i,\text{obs}}), \label{eq:likelihood}
\end{equation}
where $\mathcal{N}(x|\mu, \sigma^2)$ denotes the probability density function of a normal distribution with mean $\mu$ and variance $\sigma^2$. 

Although, we assume independent data for mass and radius, correlations can exist.
Any correlation between mass and radius can be accounted for in the likelihood function \citep{crida2018mass}. Any correlations between the data of neighbouring planets in a system, can be applied with post-processing the posterior samples \citep{dorn2018interior}.

\subsubsection{Surrogate-Based Likelihood Estimation}
As illustrated in the rapid inference phase of Fig.~\ref{fig:workflow}b, we replace the physical model in the likelihood with the fast PCK surrogate, $\hat{\mathbf{y}}(\boldsymbol{\theta})$. We adopt a pragmatic and robust approach to account for the surrogate's inherent inaccuracy, i.e., we empirically estimate a single, constant (homoscedastic) uncertainty term, $\sigma_{\text{surr}}$, calculated as the variance of the residuals on the held-out validation dataset. This term is added to the observational uncertainty to form a total variance:
\begin{equation}
\sigma^2_{i,\text{total}} = \sigma^2_{i,\text{obs}} + \sigma^2_{i,\text{surr}}. \label{eq:total_variance}
\end{equation}

This approach prevents the posterior from becoming overconfident due to surrogate model error. The final, fast-to-evaluate likelihood is then:
\begin{equation}
\mathcal{L}(\mathbf{y}_{\text{obs}}|\boldsymbol{\theta}) = \prod_{i \in \{M,R\}} \frac{1}{\sqrt{2\pi\sigma^2_{i,\text{total}}}} \exp\left(-\frac{1}{2}\frac{(\hat{y}_i(\boldsymbol{\theta}) - y_{i,\text{obs}})^2}{\sigma^2_{i,\text{total}}}\right). \label{eq:surrogate_likelihood}
\end{equation}

\subsubsection{Posterior Sampling}
We sample the posterior distribution (Eq.~\ref{eq:surrogate_likelihood}) using an Adaptive Metropolis MCMC algorithm \citep{haario2001adaptive}. For each scenario, we initialize 10 chains at random points in the admissible parameter domain and run each chain for 10,000 steps. We discard the first 2,500 steps (25\%) of each walker as a conservative burn-in phase. The use of the surrogate is critical here; a complete MCMC analysis with 100,000 total samples finishes in less than 10 minutes, whereas the same analysis using the full forward model would be computationally impractical, estimated to require several days.

\subsection{Validation Strategy} \label{sec:validation}
Our validation strategy proceeds in two complementary stages. The first stage provides a rigorous, quantitative assessment of the framework's statistical reliability through two key analyses: a direct benchmark comparison between our surrogate-based MCMC and an identical MCMC run using the computationally expensive forward model, and a large-scale coverage study to confirm that our credible intervals are correctly calibrated. The second stage addresses the framework's broad applicability by applying the entire workflow to five diverse astrophysical scenarios, ensuring it produces physically plausible results across a range of problems. \\

\section{Data} \label{sec:data}
A robust validation of our inference framework is critical, but faces a fundamental challenge: the ground-truth interior structure of any real exoplanet is unobservable. To overcome this, we use Earth as a benchmark and additionally create synthetic test cases with known properties to quantitatively assess our method's performance.

\subsection{Benchmark Comparison and Coverage Study}
To execute the validation strategy detailed in Section~\ref{sec:validation}, we conducted two in-depth analyses using a representative sub-Neptune test case (described in detail in section \ref{sec:experimental_scenarios}. For the \textit{benchmark comparison}, we performed an MCMC analysis using the computationally expensive forward model directly within the sampler, providing a direct comparison for both the runtime and the final posterior distribution against our surrogate-based results. \\

For the \textit{coverage study}, we assessed the statistical reliability of our uncertainty estimates by generating a large ensemble of 1,000 synthetic test cases. Each case was created with a known ground-truth interior structure, $\theta_{\text{true}}$, which was used as input to the expensive forward model to generate a set of ``true'' observables. From these, a final synthetic observation, $y_{\text{obs}}$, was generated for each inference run, allowing us to confirm that our credible intervals were correctly calibrated by checking if the known ground truth was captured.

For the large-scale coverage study in particular, we introduce synthetic observational uncertainty into the ground-truth test cases. For each planet's true mass ($M_{\text{true}}$) and radius ($R_{\text{true}}$), we replace them with new ``observed'' values sampled from a normal distribution centered on the true value, with standard deviations $\sigma_M$ and $\sigma_R$ representing typical observational uncertainty:
\[
M_{\text{obs}} \sim \mathcal{N}(M_{\text{true}}, \sigma_M^2)
\]
\[
R_{\text{obs}} \sim \mathcal{N}(R_{\text{true}}, \sigma_R^2)
\]

This step is crucial for a meaningful coverage analysis, also because data uncertainties are commonly considerably large.

\subsection{Experimental Scenarios} \label{sec:experimental_scenarios}
To validate our framework across a range of scientifically relevant problems, we designed five distinct experimental scenarios. The first scenario is a benchmark case based on Earth. Although Earth's mass and radius are known to high precision, we adopt observational uncertainties common for exoplanets. This setup allows us to validate the framework on a well-understood object. The second setting represents a typical super-Earth, testing the framework's performance on this common class of exoplanet. The third and fourth scenarios form a controlled experiment using a sub-Neptune target, with Setting 3 using tightly constrained observational data and Setting 4 using looser constraints. This pair is designed to demonstrate how observational precision propagates through our inference pipeline and directly impacts the breadth of the final posterior distributions. Finally, the fifth setting applies our framework to a real-world target, the well-characterized sub-Neptune TOI-270 d, allowing us to compare our inferred interior structures with results from previous studies in the literature \citep{benneke2024jwstrevealsch4co2}. The specific prior distributions and observational constraints that define these five scenarios are detailed in Table~\ref{tab:simulation_settings}.\\

\subsection{Synthetic Dataset Generation}
To generate the necessary datasets for both surrogate model development and validation, we \rev{create 1,000 simulations by drawing samples from the prior probability distributions of the input parameters using Latin Hypercube Sampling (LHS)\footnote{\rev{Latin Hypercube Sampling (LHS) is among the most popular sampling techniques and used for creating more uniform and efficient samples from a multidimensional distribution \citep{jin2005efficient}. Latin Hypercubes allow to sample variables in a controlled, stratified manner to ensure a uniform and quasi-random coverage of the sample space. It has proven to be more efficient than both random or grid sampling \citep{mckay1979comparison}.}}. For a larger number of planets with a shared prior probability distribution, the number of 1,000 simulations per planet can be reduced with a targeted, two-step process.} First, we create a large, exploratory ``master" database of $N_{\text{master}} = 5,000$ simulations by drawing samples from the prior probability distributions of the input parameters using Latin Hypercube Sampling (LHS). Second, from this master database, we extract a localized dataset of 1,000 samples relevant to a specific observational target. This is done by first defining a candidate pool of all simulations with outputs ($M$ and $R$) lying within a four-standard-deviation hyperrectangle of the target's observational values. From this subset, we perform a weighted random sampling where the probability of selecting each candidate is determined by a multivariate Gaussian distribution centered on the target observation. This targeted approach ensures the surrogate model is trained with high fidelity in the most relevant region of the parameter space, improving its accuracy where it matters most for the subsequent MCMC inference.\\

This localized dataset was then partitioned into a \textit{training dataset} of $N_{\text{train}} = 600$ samples and a \textit{held-out test dataset} of $N_{\text{test}} = 400$ samples. The training set serves a dual purpose: it is used to fit the surrogate, and its internal correlation structure defines our data-driven prior via a Gaussian copula. We adopt this data-driven, correlated prior for two primary reasons: it incorporates physically expected relationships between interior parameters, and a prior that captures the correlation structure of plausible models near the target observation can improve the efficiency of the MCMC sampling. The test set is used to evaluate the surrogate's predictive performance and to estimate its empirical error term ($\sigma_{\text{surr}}$).

\rev{The size of the training dataset of 600 samples is not a strict requirement. To assess sensitivity on the sample size, we tested a super-Earth case with a reduced training dataset. Using 100 instead of 600 samples leaves the RMS error on planetary mass essentially unchanged and increases the radius RMS error by a factor of 2.5 (from 0.008 to 0.02), still well below observational uncertainties. Even with only 10 samples, the radius RMS error rises to 0.025, which is still smaller than observational errors. This accuracy arises because PCK surrogates are designed to perform well with limited data and because the interior characterization problem is inherently smooth.}

The performance of the trained surrogate for each scenario was evaluated within the full MCMC inference framework. For the primary analysis of the five scenarios, our framework was applied directly to the observational target defined in Table~\ref{tab:simulation_settings}, where the mean values represent the data ($y_{\text{obs}}$) and the standard deviations represent the observational uncertainties ($\sigma_{\text{obs}}$).

\begin{table*}[!ht]
    \centering
    \caption{Summary of Planetary Interior Modeling Parameters for Five Simulation Settings}
    \label{tab:simulation_settings}
    \begin{tabularx}{\textwidth}{p{3.5cm} *{5}{>{\centering\arraybackslash}X}}
        \toprule
        \textbf{Parameter} & \textbf{Earth} & \textbf{Super-Earth} & \textbf{Sub-Neptune (T)} & \textbf{Sub-Neptune (L)} & \textbf{TOI-270 d} \\
        \midrule
        \multicolumn{6}{l}{\textit{\textbf{Model Setup}}} \\
        \addlinespace[0.5em]
        Mantle Model & 2 (Solid-Melt-Perplex) & 2 (Solid-Melt-Perplex) & 0 (Sotin) & 0 (Sotin) & 0 (Sotin) \\
        \midrule
        \multicolumn{6}{l}{\textit{\textbf{Compositional Priors}}} \\
        \addlinespace[0.5em]
        Mantle Fraction ($M_{\text{Mantle}}/M_{\text{Core}}$) & $\mathcal{U}(0.1, 0.9)$ & $\mathcal{U}(0.1, 0.9)$ & $\mathcal{U}(0.1, 0.9)$ & $\mathcal{U}(0.1, 0.9)$ & $\mathcal{U}(0.1, 0.9)$ \\
        Water Mass Frac. ($M_{\text{water}}/M_{\text{total}}$) & $\mathcal{U}(0, 0.05)$ & $\mathcal{U}(0, 0.05)$ & -- & -- & -- \\
        Atmosphere Mass Frac. ($M_{\text{atm}}/M_{\text{total}}$) & -- & -- & $\log X \sim \mathcal{N}(10^{-3}, 0.1))$ & $\log X \sim \mathcal{N}(10^{-3}, 0.1))$ & $\mathcal{U}(0.01, 0.1)$ \\
        Mantle Mg/Si (mass) & $0.886 \pm 0.143$ & $0.886 \pm 0.143$ & 0.886 & 0.886 & 0.886 \\
        Mantle Fe/Si (mass) & $\mathcal{U}(0, 1.69)$ & -- & -- & -- & -- \\
        \midrule
        \multicolumn{6}{l}{\textit{\textbf{Physical Priors}}} \\
        \addlinespace[0.5em]
        Age (Gyr) & -- & -- & $5.0 \pm 0.5$ & $5.0 \pm 2.0$ & $\mathcal{U}(1, 10)$ \\
        Envelope Metallicity ($Z_{\text{env}}$) & -- & -- & $0.5 \pm 0.05$ & $\mathcal{U}(0, 1)$ & $0.5 \pm 0.05$ \\
        Equilibrium Temp. (K) & $288 \pm 10$ & $1500 \pm 20$ & $700 \pm 10$ & $700 \pm 20$ & $\mathcal{U}(350, 380)$ \\
        \midrule
        \multicolumn{6}{l}{\textit{\textbf{Observational Constraints}}} \\
        \addlinespace[0.5em]
        Mass ($M_{\oplus}$) & $1.0 \pm 0.05$ & $3.0 \pm 0.3$ & $10.0 \pm 0.5$ & $10.0 \pm 1.0$ & $4.78 \pm 0.43$ \\
        Radius ($R_{\oplus}$) & $1.0 \pm 0.03$ & $1.4 \pm 0.042$ & $2.7 \pm 0.081$ & $2.7 \pm 0.135$ & $2.149 \pm 0.065$ \\
        Bulk Fe/Si (mass) & $1.69 \pm 0.353$ & $1.69 \pm 0.353$ & $1.69 \pm 0.353$ & -- & -- \\
        \bottomrule
        \addlinespace[0.5em]
        \multicolumn{6}{p{\dimexpr\textwidth-2\tabcolsep}}{\small\textbf{Notes:} Priors are given as distributions \rev{where $\log X \sim \mathcal{N}(\mu, \sigma)$ denotes a log-normal} (Gaussian) distribution with mean $\mu$ and standard deviation $\sigma$, and $\mathcal{U}(a, b)$ denotes a uniform distribution between $a$ and $b$. ``T'' and ``L'' in column headers stand for Tightly and Loosely constrained, respectively. Dashes (--) indicate that the parameter is not \rev{varied} in the specified setting. \rev{We mostly use uninformed uniform priors or priors informed by gaussian distributed observables (e.g., age, mantle Mg/Si \citep{Dorn2015}). In parts, we also vary the priors to test the impact of different prior choices, with the final selection justified on a case-by-case basis.}}  \\
    \end{tabularx}
\end{table*}

\section{Results} \label{sec:results}
\subsection{Preliminary test: Surrogate Model Validation}
\label{subsec:surrogate_validation}
Before being used for inference, the PCK surrogate model was validated on the held-out test set for each of the five experimental scenarios. This validation confirms that the surrogate is a high-fidelity replacement for the physics-based forward model. Quantitative performance metrics, shown in Table \ref{tab:nrmse_results_full}, demonstrate the surrogate's accuracy. Across all five settings, the normalized Root Mean Square Error (RMSE) is consistently below observational errors for most outputs, indicating an excellent fit. Even in the most challenging case---the loosely constrained sub-Neptune---the RMSE value for the radius is at least a factor of 3 below observational errors. We also tested the coefficient of determination ($R^2$), which is mostly above 0.99, as well as Root Mean Square Error (RMSE) and Mean Absolute Error (MAE) values, which are not shown here, which further confirm the surrogate's predictive power.



\begin{table*}[t] 
\centering
\caption{Normalized Root Mean Square Error (NRMSE) of the surrogate model for each experimental setting. The NRMSE is calculated by normalizing the RMSE by the mean of the true values in the test set.}
\label{tab:nrmse_results_full}
\begin{tabular}{lccc}
\toprule
\textbf{Experimental Setting} & \textbf{NRMSE Mass ($M_\oplus$)} & \textbf{NRMSE Radius ($R_\oplus$)} & \textbf{NRMSE Fe/Si$_{\text{bulk}}$} \\
\midrule
1: Earth             & 0.34\% & 0.10\% & 0.31\% \\
2: Super-Earth       & 1.15\% & 0.24\% & 0.05\% \\
3: Sub-Neptune (T)   & 0.37\% & 0.10\% & 0.48\% \\
4: Sub-Neptune (L)   & 1.16\% & 1.76\% & N/A    \\
5: TOI-270d          & 0.43\% & 0.25\% & N/A    \\
\bottomrule
\end{tabular}
\end{table*}
Crucially, the magnitude of this model error, as quantified by the RMSE, is typically an order of magnitude smaller than the observational uncertainties associated with exoplanet characterization \citep{Weiss_2014}. Consequently, the surrogate model error is expected to have a minimal impact on the final posterior distributions.

\begin{figure*}[htbp]
    \centering
    \begin{subfigure}{\textwidth}
        \centering
        \includegraphics[
            width=1.\textwidth, 
            page=1, 
            trim={1.2cm 0cm 0cm 1cm}, 
            clip
        ]{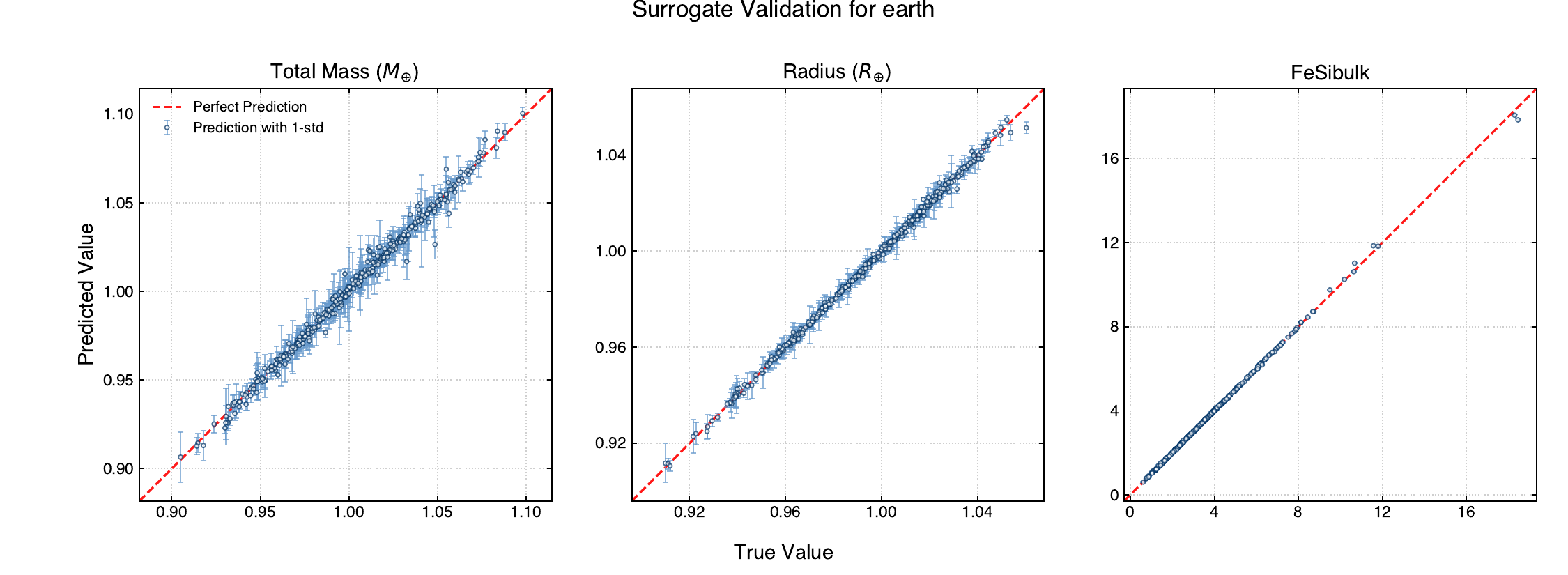}
    \end{subfigure}\\[0.1cm]
    
    \begin{subfigure}{\textwidth}
        \centering
        \includegraphics[
            width=0.7\textwidth, 
            page=1, 
            trim={0cm 0cm 0cm 1cm}, 
            clip
        ]{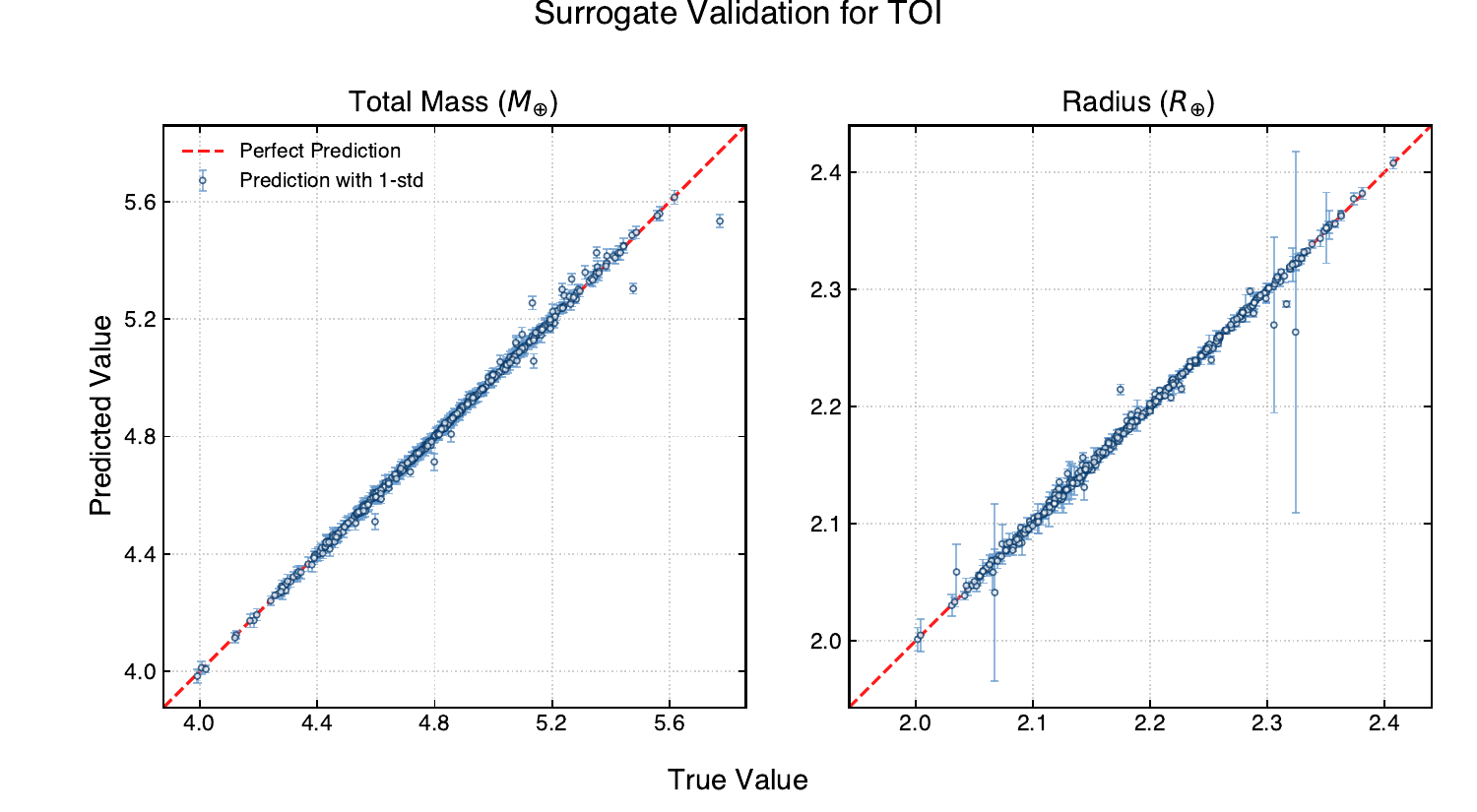}
    \end{subfigure}
    
    \caption{Validation of the PCK surrogate's predictive performance on the held-out test set for two representative cases: (a) the Earth (Setting 1) and (b) the TOI-270 d (Setting 5). The surrogate's mean prediction (blue dot) shows agreement with the true forward model outputs (red dotted line). 
    Results for the remaining experimental settings are shown in Appendix~\ref{app:PCK_val_plots}.}
    \label{fig:surrogate_prediction}
\end{figure*}

This performance is visualized in the residual plots shown in Figure \ref{fig:surrogate_prediction}. For both the Earth (Setting 1) and TOI-270 d (Setting 5) scenarios, the residuals are tightly scattered around zero with only minor outliers, confirming that the surrogate is generally unbiased and accurate. 

The validation plots for the remaining three scenarios, provided in Appendix~\ref{app:PCK_val_plots}, show similar overall performance.

\subsection{Stage 1: Statistical Reliability and Performance}
This first validation stage provides a rigorous, quantitative assessment of the framework's statistical integrity and computational performance. We conduct two complementary analyses: a direct benchmark against the full forward model and a large-scale coverage study to confirm that our credible intervals are correctly calibrated.

\subsubsection{Benchmark Comparison: Surrogate vs. Full Forward Model}
To validate our surrogate-based inference framework, we performed a direct comparison between our PCK surrogate approach and an MCMC run using the full forward model for the Sub-Neptune (T) (Setting 3) scenario. Both analyses used identical observational constraints ($M = 10.0 \pm 0.5 M_{\oplus}$, $R = 2.7 \pm 0.081 R_{\oplus}$) and prior distributions.

The corner plots showing the posterior distributions from both approaches are available in the Appendix~\ref{app:benchmark_plots}. The inferred parameters are summarized below:

\begin{center}
\begin{tabular*}{\columnwidth}{l @{\extracolsep{\fill}} r r}
    \toprule
    \textbf{Parameter} & \textbf{Surrogate} & \textbf{Benchmark} \\
    \midrule
    $M_{\text{atm}}$ [$M_{\oplus}$] & $0.52^{+0.15}_{-0.13}$ & $0.44^{+0.16}_{-0.13}$ \\[1ex]
    $M_{\text{mantle}}$ [$M_{\oplus}$] & $6.70^{+0.89}_{-0.78}$ & $6.92^{+0.93}_{-0.86}$ \\[1ex]
    $M_{\text{core}}$ [$M_{\oplus}$] & $2.80^{+0.67}_{-0.76}$ & $2.66^{+0.72}_{-0.77}$ \\[1ex]
    $Z_{\text{env}}$ & $0.50^{+0.05}_{-0.05}$ & $0.48^{+0.05}_{-0.05}$ \\
    \bottomrule
\end{tabular*}
\end{center}

The posterior means agree within $0.6\sigma$ for all parameters, and the credible intervals are broadly consistent. This comparison demonstrates that our surrogate approach preserves statistical integrity while achieving substantial computational savings.

\subsubsection{Coverage Analysis}
This stage provides a rigorous, quantitative test of the statistical reliability of our framework's uncertainty estimates. We ran our MCMC pipeline on the large ensemble of 1,000 synthetic planets for which the ground-truth interior parameters, $\theta_{\text{true}}$, were known.

For each run, we checked whether the known ground truth was captured within the 68\% and 95\% credible intervals of the inferred posterior. The results, summarized in Table~\ref{tab:coverage_results}, show coverage probabilities that are close to the nominal values; the 68\% and 95\% credible intervals captured the ground truth in 65–72\% and 93–96\% of cases, respectively.

\begin{table}[htbp!]
    \centering
    \caption{Coverage probabilities for the Sub-Neptune (T) scenario, calculated from 1000 synthetic test cases. The values represent the fraction of runs in which the known ground truth was captured by the posterior's credible interval.}
    \label{tab:coverage_results}
    \begin{tabular}{lcc}
        \toprule
        Parameter & 68\% Coverage & 95\% Coverage \\
        \midrule
        Atmospheric mass & 65\% & 93\% \\
        Mantle mass & 69\% & 93\% \\
        Core mass & 72\% & 96\% \\
        \bottomrule
    \end{tabular}
\end{table}

\subsubsection{Computational Performance and Speedup}
\label{subsec:speedup}
To demonstrate the practical advantage of the surrogate-based approach, we quantified the computational speedup achieved by replacing the full physics-based forward model with the PCK surrogate. We performed a direct comparison for the tightly constrained Sub-Neptune case (T) (Setting 3), as its complexity provides a representative benchmark.

Two identical MCMC inferences were run on a single CPU core (Apple M4 processor with 16 GB RAM): one using the original forward model and the other using the trained PCK surrogate. Both analyses collected 100.000 posterior samples to ensure proper convergence. The benchmark run using the full forward model required approximately 42 hours to complete, while our surrogate-based inference finished the identical analysis in just 8 minutes.

For this specific problem, we observe computational speedup factors of $\sim 300 - 2000$, depending on the number of parallel chains (10-1000), reducing inference times by 2-3 orders of magnitude. The dramatic reduction in runtime transforms what is computationally feasible in exoplanet interior studies. Most significantly, this efficiency enables extensive validation exercises that would otherwise be computationally prohibitive—such as our 1,000-run coverage analysis—and makes rigorous statistical inference tractable for population-scale studies involving hundreds or thousands of exoplanets.

\subsection{Stage 2: Broad Applicability on Synthetic Targets}
With the framework's statistical reliability established in Stage 1, this second stage demonstrates its practical application to the five synthetic targets. The goal is to ensure the pipeline produces physically plausible and well-behaved posterior distributions. To verify this, we inspect the physical plausibility of the outcomes, check that expected physical relationships (such as the anti-correlation between core and mantle masses) are preserved, and confirm that posteriors respond appropriately to the strength of the observational data.

For these five scenarios, we first verified the MCMC sampler's performance. Sampler acceptance rates ranged from 16–25\%, aligning with the theoretical optimum of $\approx23.4\%$ and indicating a reliable sampling process \citep{gelman2013bayesian, roberts2001optimal}. 

For each scenario, the MCMC was run using the synthetic observation defined in the ``observational constraints" section of the settings table (see Table~\ref{tab:simulation_settings}).

\begin{figure*}[t]
    \centering
    \begin{subfigure}{0.48\textwidth}
        \centering
        \includegraphics[
            width=\linewidth,
            page=2, 
            trim={0cm 0cm 0cm 0cm}, 
            clip
        ]{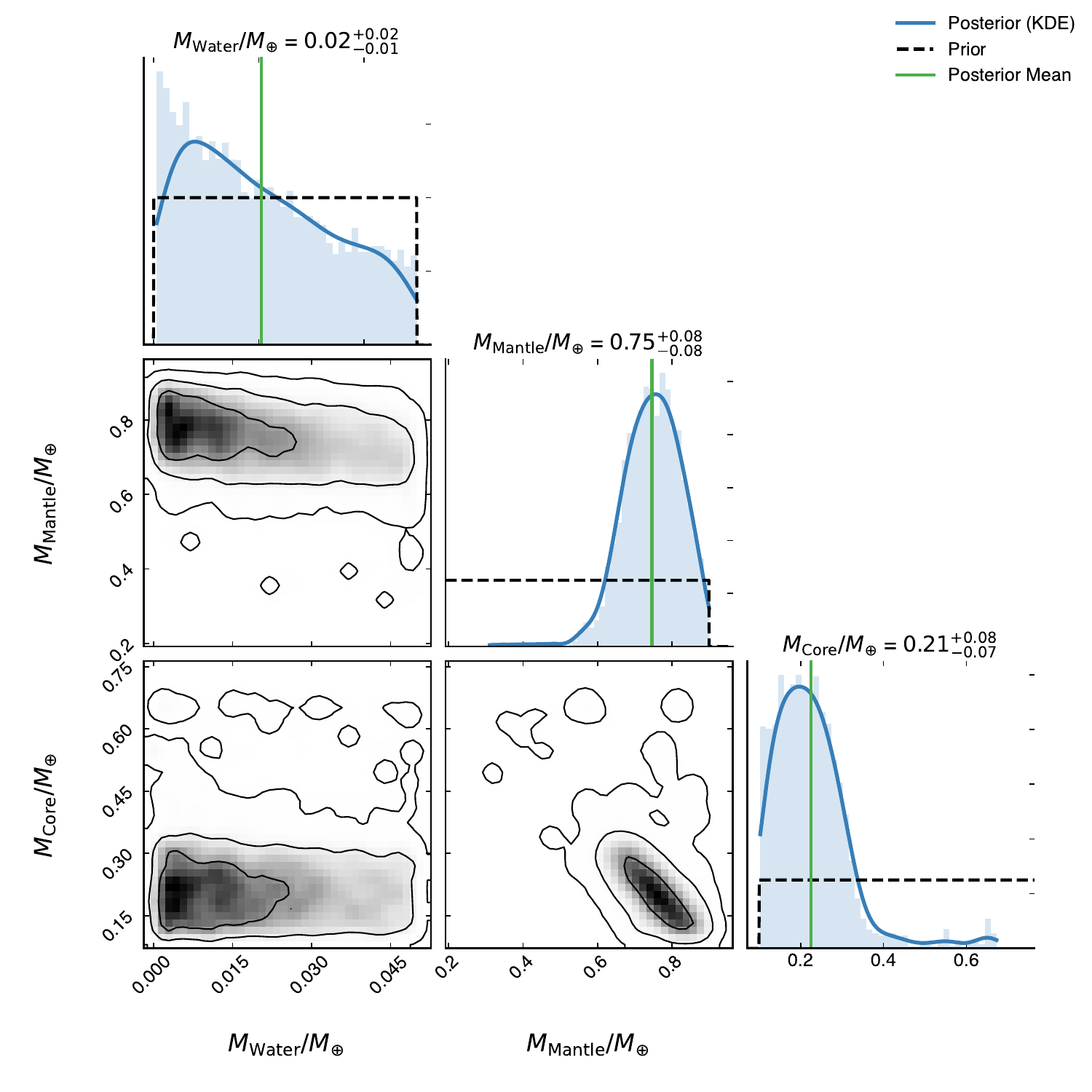}
        \caption{Earth (Setting 1)}
        \label{fig:corner_plot_earth}
    \end{subfigure}
    \hfill 
    \begin{subfigure}{0.48\textwidth}
        \centering
        \includegraphics[
            width=\linewidth,
            page=2, 
            trim={0cm 0cm 0cm 0cm}, 
            clip
        ]{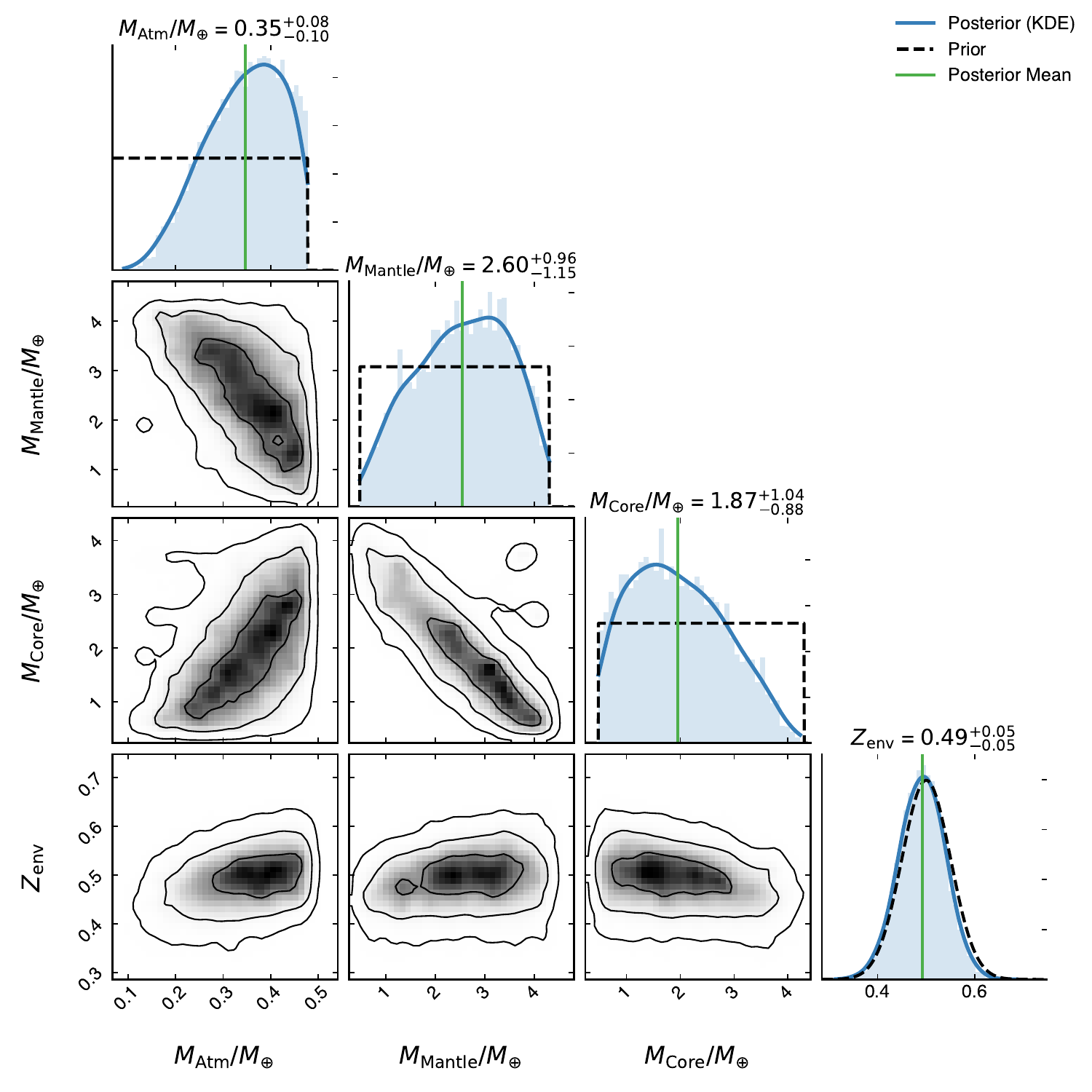}
        \caption{TOI-270 d (Setting 5)}
        \label{fig:corner_plot_TOI}
    \end{subfigure}
    
    \caption{Posterior distributions for the interior structure parameters inferred via MCMC, shown for two representative cases: (a) the constrained Earth case (Setting 1) and (b) the TOI-270 d exoplanet (Setting 5). The results demonstrate the framework's ability to produce well-behaved posteriors that reflect the strength of the observational constraints. Corner plots for all other scenarios are available in Appendix~\ref{app:corner_plots}.}
    \label{fig:mcmc_corner_plots}
\end{figure*}

Figure~\ref{fig:mcmc_corner_plots} shows representative results for two scenarios: the Earth case (Setting 1) and the TOI-270 d case (Setting 5). These corner plots demonstrate the framework's ability to produce well-behaved posteriors that appropriately reflect the observational constraints.

\subsubsection{Earth}
As a primary validation of our framework, we tested it on a synthetic Earth-like planet (Setting 1), for which planet Earth offers independent interior constraints from geophysical data. The resulting posterior distributions for the interior parameters are shown in Figure~\ref{fig:corner_plot_earth}. Our analysis infers a core mass of $0.21^{+0.08}_{-0.07}$~M$_{\oplus}$ and a mantle mass of $0.75^{+0.08}_{-0.08}$~M$_{\oplus}$.

We compare these results to the known composition of Earth, which has a core mass of approximately 0.32~M$_{\oplus}$ and a mantle mass of about 0.68~M$_{\oplus}$. Our inferred mean for the mantle mass lies approximately \textbf{1.0$\sigma$} from the true value, while the mean core mass is about \textbf{1.4$\sigma$} away. This deviation is expected and arises directly from the observational uncertainties imposed on the input data (5\% on mass and 3\% on radius). Because the inference framework must account for this uncertainty, it identifies a range of plausible interior structures rather than a single exact solution. This effect is compounded by the inherent degeneracy of the problem, where different core-mantle configurations can produce similar total masses and radii. Consequently, the resulting posterior distributions are broader than the single ground-truth values but contain them.

\subsubsection{Application to TOI-270 d}
We applied our validated framework to the sub-Neptune TOI-270 d (Setting 5) \citep{gunther2019super} to demonstrate its practical utility. Similar to \cite{benneke2024jwstrevealsch4co2}, we make use of the constraint on atmospheric metallicity from transmission spectroscopy with $Z_{env} = 0.5 \pm 0.05$. The inferred interior structure parameters are presented in Figure~\ref{fig:corner_plot_TOI}. Our analysis yields a core mass of $1.87^{+1.04}_{-0.88}$~M$_{\oplus}$, a mantle mass of $2.60^{+0.96}_{-1.15}$~M$_{\oplus}$, and an atmospheric mass of $0.35^{+0.08}_{-0.10}$~M$_{\oplus}$.

To place these findings in context, we compare them to the results of \cite{benneke2024jwstrevealsch4co2}. Their interior model, which assumes a Earth-like rocky interior inferred the combined mass of core  and mantle to be $4.28^{+0.46}_{-0.47}$~M$_{\oplus}$. To make a direct comparison, we sum our inferred core and mantle masses to find a total rocky mass of \textbf{$4.47^{+1.42}_{-1.44}$~M$_{\oplus}$}. Our result is in agreement with their estimate, demonstrating the consistency of our framework with interior characterizations which are additionally constrained by atmospheric retrievals. The higher uncertainty in our estimates stems from the fact that we do not fix an Earth-like rocky interior.

\subsubsection{Framework Validation Across Constraint Strengths}
The results clearly demonstrate how the framework responds to different levels of observational constraint strength. Comparing scenarios with varying constraint tightness reveals that the posteriors correctly capture the increased model degeneracy when fewer or less precise observations are available. For instance, the posteriors for the sub-Neptune scenario (L) (Setting 4) showed significantly broader distributions and more pronounced parameter correlations than its tightly constrained counterpart, the Sub-Neptune (T) (Setting 3) scenario, with posterior standard deviations that were substantially larger in the loosely constrained case.
Corner plots for the remaining scenarios (Super-Earth, Sub-Neptune tight, and Sub-Neptune loose) are provided for completeness in Appendix~\ref{app:corner_plots}.

\section{Discussion} \label{sec:discussion}
Our surrogate-accelerated Bayesian inference framework represents a significant advancement in exoplanet interior characterization, addressing the fundamental computational bottleneck that has limited the scope and statistical rigor of such analyses. The results demonstrate that polynomial chaos-Kriging surrogates can effectively replace expensive physics-based forward models while maintaining high fidelity and enabling rigorous uncertainty quantification.

\subsection{Computational Efficiency and Scalability}
The computational speedup achieved by our framework---reducing inference times from days to minutes---transforms what is computationally feasible in exoplanet interior studies. This efficiency gain enables several previously impractical research directions. First, it makes large-scale population studies tractable, allowing researchers to systematically characterize hundreds or thousands of exoplanets with rigorous statistical methods. Second, it facilitates comprehensive sensitivity analyses and model comparison studies that require numerous MCMC runs. Third, the rapid inference capability enables real-time analysis workflows, where new observational data can be quickly incorporated to update interior structure estimates.

Our approach is highly scalable. The fast surrogate model enables the deployment of numerous parallel MCMC chains, shifting the computational bottleneck from model evaluation to sampling. A higher number of MCMC chains can further accelerate convergence and facilitate a more comprehensive exploration of the posterior distribution.
This scalability is particularly valuable given the exponentially growing catalogue of confirmed exoplanets. Traditional methods that require days of computation per planet become prohibitive for population-scale analyses, whereas our framework maintains statistical rigour while achieving the throughput necessary for systematic studies across diverse planetary populations.

\subsection{Framework Flexibility and Adaptability}
A key advantage of our approach over alternative methods like conditional Invertible Neural Networks (cINNs) or Mixture Density Networks (MDNs) is its adaptability to evolving physical models. While neural network approaches require extensive retraining with millions of evaluations when the underlying physics changes, our surrogate can be retrained with just a few hundred model evaluations. This efficiency stems from several fundamental differences: PCK models have fewer hyperparameters to optimize compared to deep neural networks, the underlying optimization problem is largely convex (particularly for the polynomial basis fitting), and the hybrid architecture leverages well-established statistical methods (polynomial regression and Gaussian processes) rather than complex non-linear transformations that require extensive exploration of parameter space. This flexibility is essential in the rapidly evolving field of exoplanet interior modelling, where new physical processes, thermodynamic data, and compositional constraints are regularly incorporated.

The modular design of our framework also facilitates methodological improvements. For instance, the surrogate component can be enhanced with more sophisticated uncertainty quantification techniques, without requiring fundamental changes to the overall architecture.

\subsection{Implications for Exoplanet Science}
Our results for TOI-270~d demonstrate practical applicability to real exoplanets, with inferred interior parameters consistent with previous detailed studies. This consistency provides confidence that our framework can contribute meaningful scientific insights while reducing computational requirements.

The framework's ability to handle diverse planetary scenarios---from Earth-like planets to sub-Neptunes with substantial atmospheres---indicates broad applicability across the exoplanet population.

\subsection{Future Developments}
Several enhancements could further improve the framework's capabilities. Implementing input-dependent uncertainty quantification for the surrogate model could provide more accurate error propagation. The framework could be extended to propagate uncertainties inherent in the original forward model itself, providing a more comprehensive assessment of the overall uncertainty. In addition, the training of the surrogate can be made adaptive such that expensive forward model evaluations are only  added where the surrogate uncertainty is high and the posterior has significant probability mass \citep{li2014adaptive}. Furthermore, incorporating additional observational constraints, such as atmospheric composition data or stellar abundances, could help break parameter degeneracies. 

\section{Conclusion} \label{sec:conclusion}
We have developed and validated a surrogate-accelerated Bayesian inference framework that addresses the computational bottleneck in exoplanet interior characterization. Our approach replaces expensive physics-based forward models with fast Polynomial Chaos-Kriging surrogates. In practical terms, it replaces the computational burden from repeatedly evaluating expensive forward models with the cost of generating a few hundred models for surrogate training \rev{for a single planet}. For our employed interior forward model this results in a speed-up of 2-3 orders of magnitude. At the same time, we maintain statistical rigor and physical fidelity.

Key validation results demonstrate: (1) surrogate accuracy with $R^2 > 0.99$ for most scenarios—from Earth-like worlds to sub-Neptunes—with RMSE values typically an order of magnitude smaller than observational uncertainties, ensuring that surrogate error does not significantly impact posterior distributions; (2) statistical reliability tested through 1000 synthetic test cases with coverage probabilities of 65–72\% and 93–96\% for the nominal 68\% and 95\% intervals, respectively; and (3) successful application to real targets, with TOI-270 d results consistent with previous detailed studies.

The framework also presents opportunities to rigorously use additional constraints on atmospheric composition from missions like JWST, Ariel, or ELT. Connecting compositional information from the upper atmosphere to a planet’s deeper interior requires coupled atmosphere–interior models, which can significantly increase computational cost. Our approach alleviates this challenge by enabling the use of comprehensive models within inference frameworks, since the number of required forward model evaluations is limited to only a few hundred to thousand runs.

The framework's computational efficiency also enables rigorous population-scale studies that were previously impractical. Its reliance on only a few hundred physics-based forward models ensures adaptability to ongoing model developments, maintaining relevance as physical models evolve. This makes our approach particularly well suited for model comparisons in inference problems, where the community is often interested in quantifying the impact of different model assumptions on the results. By requiring only a limited set of physics-based models, our method substantially reduces computational demands and makes it feasible to include even time-intensive models in inference analyses.
This represents a significant advancement toward systematic characterization of exoplanet interior diversity across the growing catalog of confirmed planets, with particular relevance for upcoming large-scale surveys such as the PLATO mission, as well as data from Kepler, TESS and CHEOPS.

\section*{Acknowledgements}


\bibliography{references}{}
\bibliographystyle{aa}

\appendix
\section{Benchmark Comparison: Corner Plots}
\label{app:benchmark_plots}

\begin{figure*}[htbp]
    \centering
    \begin{subfigure}{0.48\textwidth}
        \centering
        \includegraphics[
            width=\textwidth,
            page=2,
            trim={0cm 0cm 0cm 0cm}, 
            clip
        ]{figures/sub_nep_tight.pdf}
        \caption{Surrogate-based MCMC\\(8 minutes)}
        \label{fig:benchmark_surrogate_side}
    \end{subfigure}
    \hfill
    \begin{subfigure}{0.48\textwidth}
        \centering
        \includegraphics[
            width=\textwidth,
            page=1,
            trim={0cm 0cm 0cm 0cm},
            clip
        ]{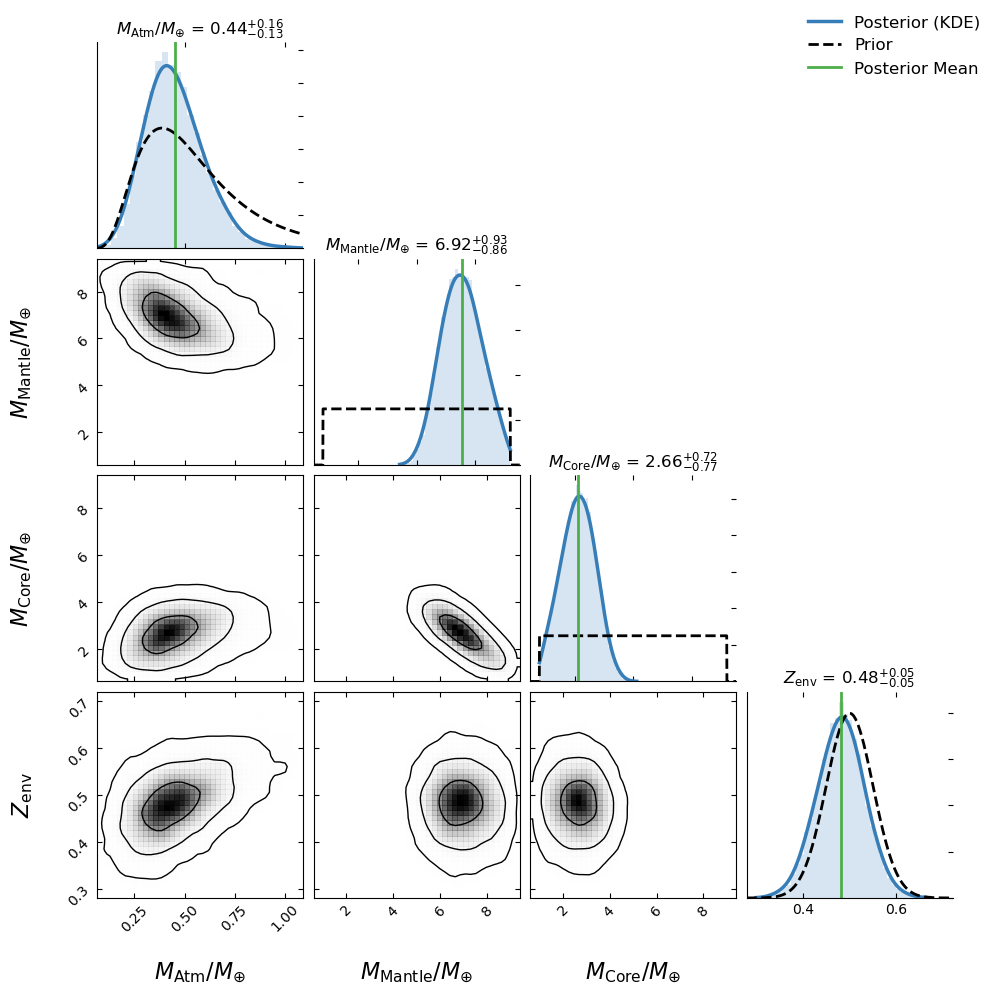}
        \caption{Full Forward Model MCMC\\(42 hours)}
        \label{fig:benchmark_physics_side}
    \end{subfigure}
    
    \caption{Side-by-side comparison of posterior distributions for the Sub-Neptune (T) scenario obtained using (a) our surrogate-based approach and (b) the full forward model. The computational times highlight the dramatic efficiency gain achieved by our framework.}
    \label{fig:benchmark_comparison_side}
\end{figure*}

\clearpage

\section{PCK model validations}
\label{app:PCK_val_plots}

\begin{figure*}[htbp!]
    \centering
    
    \begin{subfigure}{\textwidth}
        \centering
        \includegraphics[width=0.85\textwidth, page=1, trim={0cm 0cm 0cm 1cm}, clip]{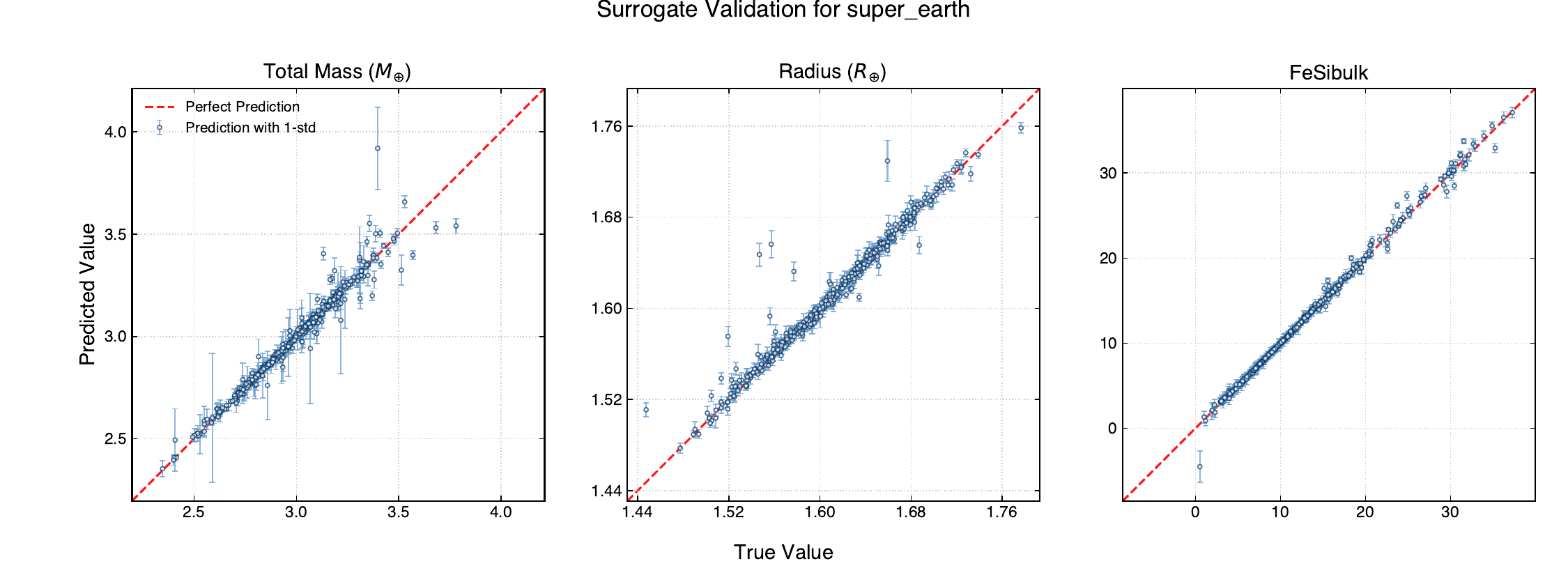}
        \caption{Super-Earth (Setting 2)}
        \label{fig:pck_val_super_earth} 
    \end{subfigure}
    
    \vspace{0.5cm} 
    
    \begin{subfigure}{\textwidth}
        \centering
        \includegraphics[width=0.85\textwidth, page=1, trim={0cm 0cm 0cm 1cm}, clip]{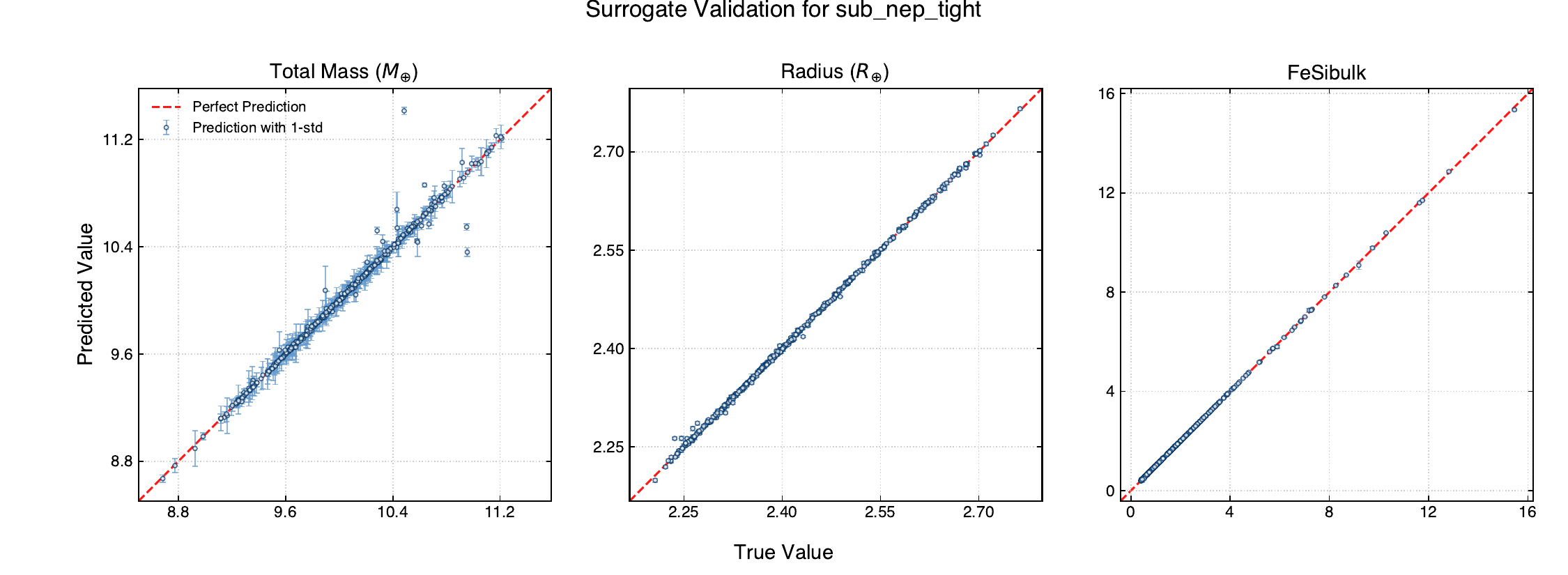}
        \caption{Sub-Neptune (T) (Setting 3)}
        \label{fig:pck_val_sub_nep_tight} 
    \end{subfigure}
    
    \vspace{0.5cm}
    
    \begin{subfigure}{\textwidth}
        \centering
        \includegraphics[width=0.55\textwidth, page=1, trim={0cm 0cm 0cm 1cm}, clip]{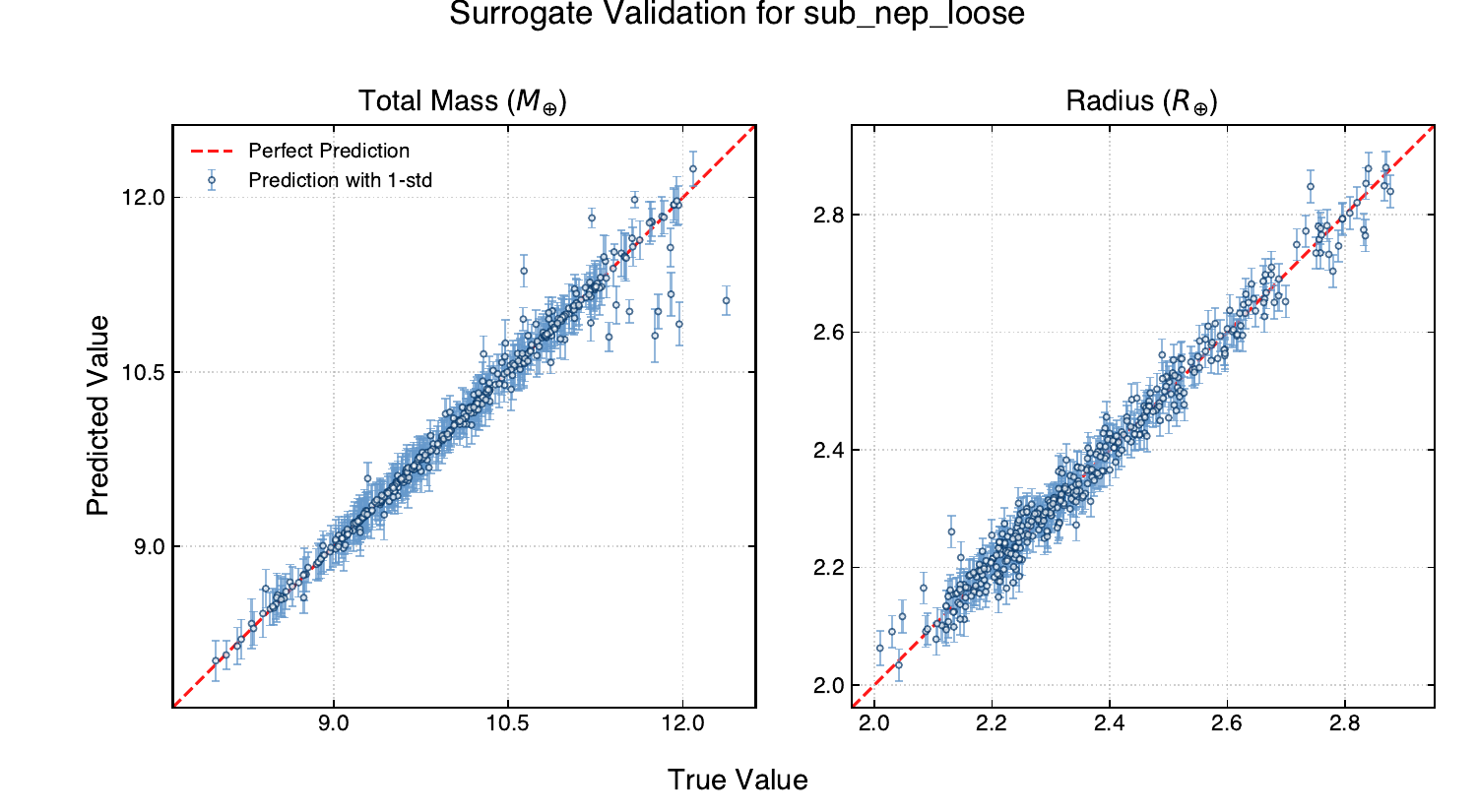}
        \caption{Sub-Neptune (L) (Setting 4)}
        \label{fig:pck_val_sub_nep_loose} 
    \end{subfigure}

    \caption{Residual plots showing the predictive performance of the PCK surrogate on the held-out test sets for the Super-Earth, tightly constrained Sub-Neptune, and loosely constrained Sub-Neptune scenarios. The residuals (Predicted - True) are tightly scattered around zero, confirming the surrogate's high fidelity.}
    \label{fig:appendix_pck_validation}
\end{figure*}

\clearpage 

\section{MCMC Convergence Diagnostics}
\label{app:convergence} 
To ensure the reliability of our MCMC analysis, we assessed the convergence and mixing of the sampler for all scenarios. Most scenarios demonstrated excellent convergence (Gelman-Rubin statistic $\hat{R} < 1.01$). The Earth scenario showed slightly elevated $\hat{R}$ values, which however, became low again when running the chains longer.
In all cases, we observed a healthy acceptance rate ($\sim 20\%$) that confirms the sampler was exploring the space effectively.

\begin{table}[htbp]
\centering
\caption{MCMC Acceptance Rates}
\label{tab:convergence_stats}
\begin{tabular}{lc}
    \toprule
    Scenario & Acceptance Rate (Mean $\pm$ Std) \\
    \midrule
    Setting 1: Earth & $0.20 \pm 0.04$ \\
    Setting 2: Super-Earth & $0.16 \pm 0.02$ \\
    Setting 3: Sub-Neptune (T) & $0.25 \pm 0.02$ \\
    Setting 4: Sub-Neptune (L) & $0.24 \pm 0.01$ \\
    Setting 5: TOI-270 d & $0.23 \pm 0.01$ \\
    \bottomrule
\end{tabular}
\end{table}

\clearpage

\section{Corner Plots for Remaining Scenarios}
\label{app:corner_plots}

This section provides the posterior distributions for the interior structure parameters for the three scenarios not shown in the main text.

\begin{figure*}[htbp]
    \centering
    
    \begin{subfigure}{0.48\textwidth}
        \centering
        \includegraphics[width=\textwidth, page=2]{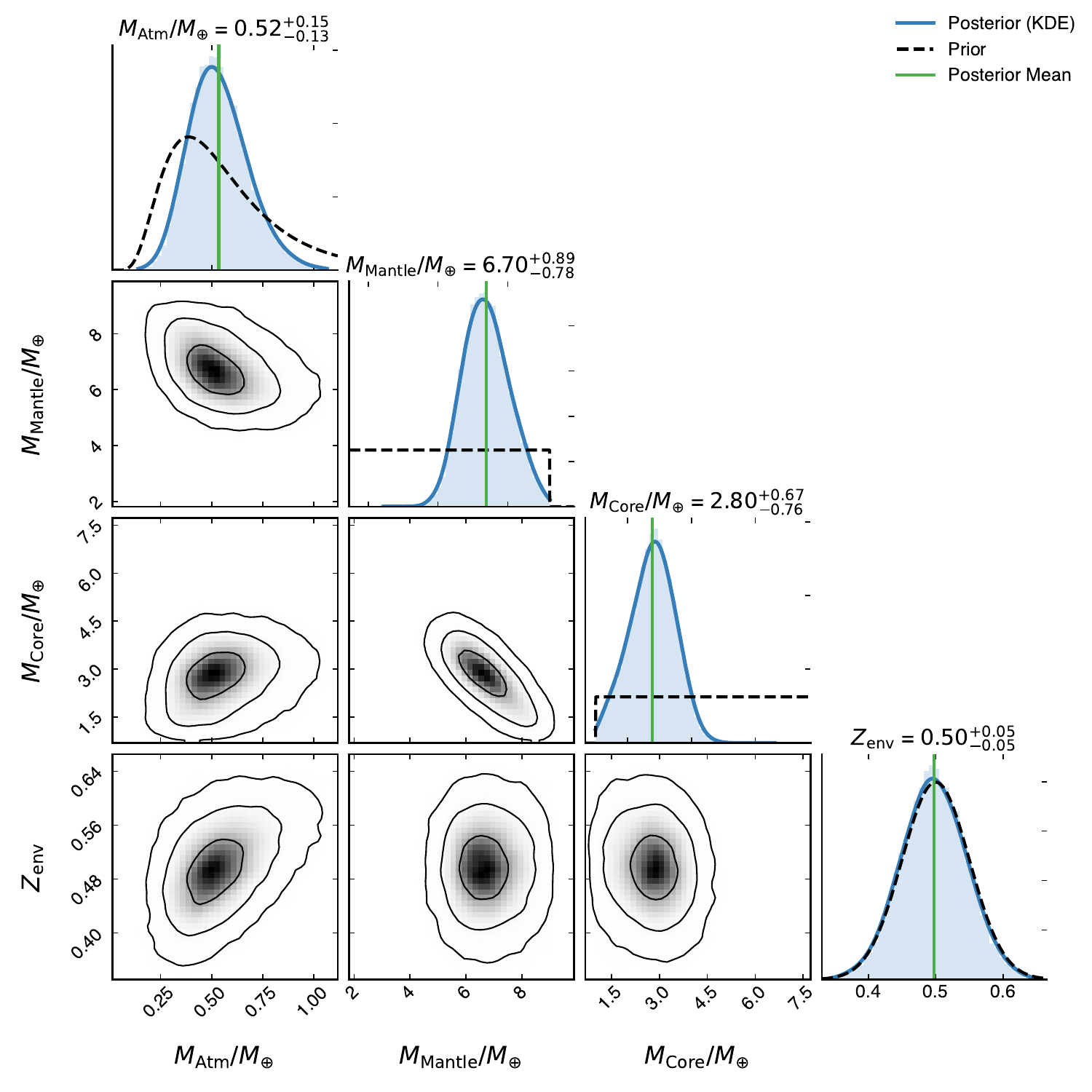}
        \caption{Sub-Neptune (T) (Setting 3)}
        \label{fig:corner_sub_neptune_tight}
    \end{subfigure}
    \hfill 
    \begin{subfigure}{0.48\textwidth}
        \centering
        \includegraphics[width=\textwidth, page=2]{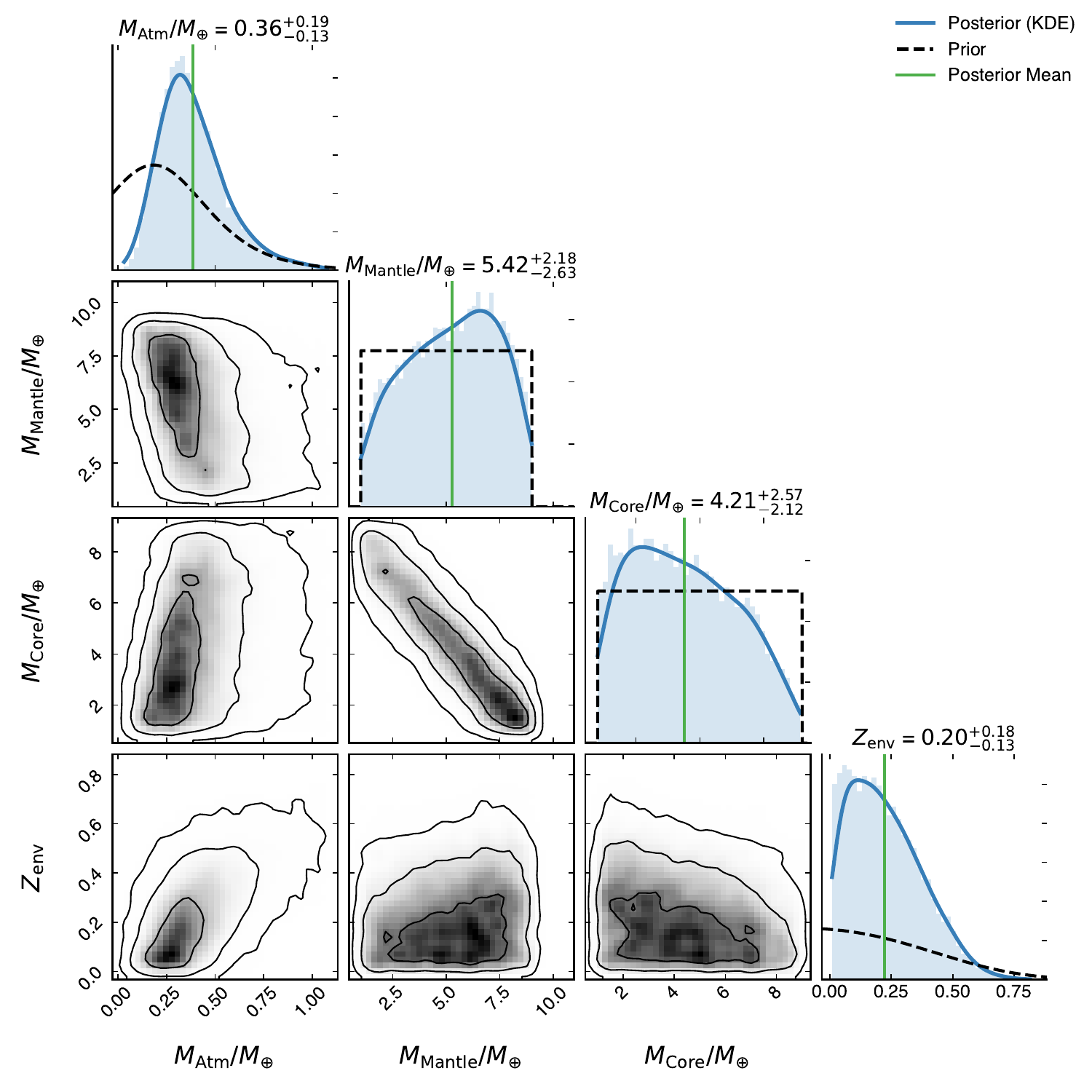}
        \caption{Sub-Neptune (L) (Setting 4)}
        \label{fig:corner_sub_neptune_loose}
    \end{subfigure}

    \vspace{0.2cm} 

    \begin{subfigure}{0.5\textwidth} 
        \centering
        \includegraphics[width=\textwidth, page=2]{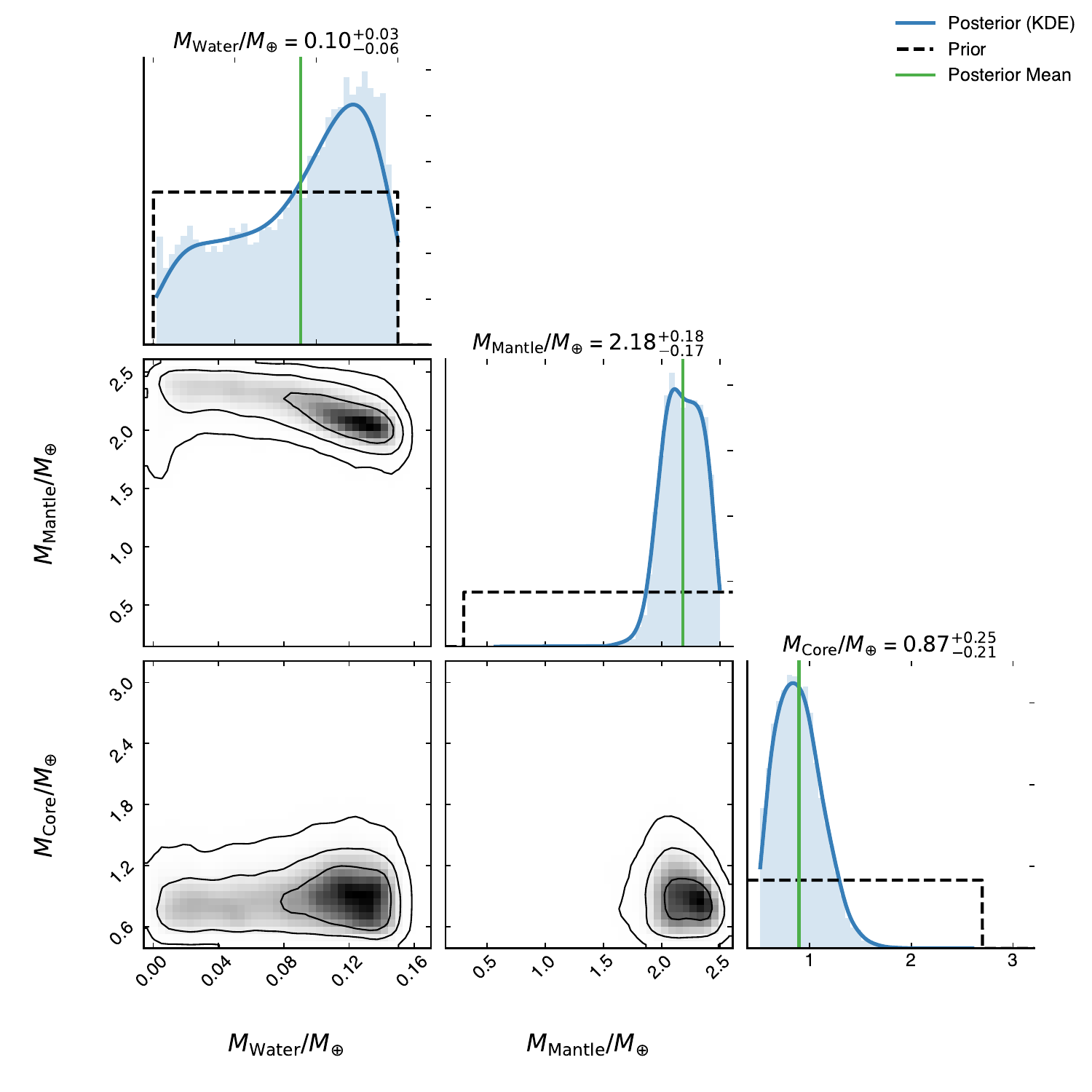}
        \caption{Super-Earth (Setting 2)}
        \label{fig:corner_super_earth}
    \end{subfigure}

    \caption{Posterior distributions for the interior structure parameters of the remaining scenarios. The top row compares the (a) tightly constrained and (b) loosely constrained Sub-Neptune cases, showing the impact of precise observational data. The bottom panel shows the (c) Super-Earth scenario.}
    \label{fig:appendix_corners_grid}
\end{figure*}



\end{document}